\documentclass[12pt,a4j]{article}
\usepackage[dvips]{color,graphicx}
\usepackage{amsmath,enumerate, amsfonts, ulem, url}
\usepackage{natbib}
\begin{document}
\baselineskip 18pt
\begin{center}
{\LARGE\textbf{Selection of variables and decision boundaries for functional data via bi-level selection}}
\end{center}
\begin{center}
{\large Hidetoshi Matsui}
\end{center}

\begin{center}
\begin{minipage}{14cm}
{
\begin{center}
{\it {\footnotesize 
The Center for Data Science Education and Research, Shiga University \\
1-1-1, Banba, Hikone, Shiga 522-8522, Japan. \\
}}

\vspace{2mm}

{\small hmatsui@biwako.shiga-u.ac.jp}
\end{center}
%-----------------------------------------------
\vspace{1mm} 

{\small {\bf Abstract:} 
	Sparsity-inducing penalties are useful tools for variable selection and they are also effective for regression settings where the data are functions.  
	We consider the problem of selecting not only variables but also decision boundaries in logistic regression models for functional data, using the sparse regularization.  
	The functional logistic regression model is estimated by the framework of the penalized likelihood method with the sparse group lasso-type penalty, and then tuning parameters are selected using the model selection criterion.  
	The effectiveness of the proposed method is investigated through %simulation studies and 
	real data analysis.
}

\vspace{3mm}

}
\end{minipage}
\end{center}

\section{Introduction}
Variable selection is one of the most important issues in regression analysis and several methods have been proposed for the accurate and effective selection of appropriate variables \citep[see, e.g.,][]{KoKi2008}. 
For such problems, sparse regularization that estimates the model with $L_1$-type penalties provides a unified approach for estimating and selecting variables, and for this reason they are broadly applied in several fields \citep{Buva2011, HaTiWa2015}. 
In this paper, we consider the problem of selecting not only variables but also decision boundaries which affect the classification problem, by applying the sparse regularization to logistic regression models when the data to be classified are measured repeatedly over time. 

The logistic regression model is a useful tool for classifying data, and it does so by providing posterior probabilities which place the data in the appropriate group \citep{McNe1989}. 
Logistic regression modeling that use the sparse regularization have been investigated as generalized linear models in \cite{PaHa2007} and
\cite{KrCaFi_etal2005}, and \cite{FrHaTi2010a} applied $L_1$-type penalties to the multinomial or multiclass logistic regression model that classifies data into three or more groups as natural extensions of the binomial logistic regression models. 
More recently, \cite{ViHa2014} applied the sparse group lasso-type penalty \citep{SiFrHa_etal2013} to the logistic regression model.  
The sparse group lasso is one of the bi-level selection techniques \citep{BrHu2009, Ma2015} that select variables in both group and individual levels.
Therefore, it can be seen as a composition of the lasso \citep{Ti1996} and the group lasso \citep{YuLi2006}.  
%On the other hand, there are also multiple parameters in each variable of the multinomial logistic regression model...

Functional data analysis (FDA), which is established by Ramsay and Silverman (2005), is a useful method for effectively analyzing repeatedly measured data, and it has received considerable attentions in various fields \citep{RaSi2002, HoKo2012}. 
The basic idea behind FDA is to express repeated measurement data for each individual as a smooth function and then to draw information from the collection of these functions. 
FDA includes extensions of traditional methods, and in particular there are many works on regression models.  
For logistic regression models for functional data, there are various works in \cite{AgEs2008}, \cite{AgAgEs_etal2013}, and \cite{EsAgVa2007}.  
Furthermore, the problem of variable selection for functional regression models using $L_1$-type regularization is considered in \cite{FeHaVi2010}, \cite{AnFeVi2011}, \cite{MaKo2011}, \cite{ZhOgRe2012}, \cite{GeMaSt2013}, and \cite{MiLiRo2013}. 
%2014-02-14 revise2-------
However, these works do not include the multiclass logistic regression model.  
For this model, we may fail to select functional variables when we use existing types of penalties, since it has multiple coefficients for multiple decision boundaries.  
In order to solve this problem, \cite{Ma2014} proposed two types of penalties for selecting variables and decision boundaries respectively.  

In this paper we apply the bi-level selection technique to the functional logistic regression model in order to select variables and decision boundaries simultaneously.  
Time course observations are smoothed by using basis expansions, and then parameters included in the functional logistic regression model are estimated by the sparse regularization with the sparse group lasso-type penalty.  
We apply the blockwise descent algorithm derived by \cite{SiFrHa_etal2013} for estimating the coefficient parameters.
Values of tuning parameters in the penalty function are selected by a model selection criterion.  
%Through simulation studies, we show that the proposed method can select both variables and decision boundaries. 
%Furthermore, we report the result of the real data analysis.
The effectiveness of the proposed method is investigated through the real data analysis.  

This paper is organized as follows.  
In Section 2, we introduce the details of the logistic regression model for functional data and some preparations for estimation of the model.
Section 3 provides the method for estimating coefficient parameters and for selecting tuning parameters.  
%Simulation results are shown in Section 4 and then 
The results of real data analysis are discussed in Section 4.  
Finally we conclude the article with some discussions in Section 5.  
\section{Multiclass logistic regression model for functional data}
Suppose we have $n$ sets of functional data and class labels 
$\{(x_{i}(t), g_i);$ $i=1,\ldots, n\}$, where 
$x_i(t) = (x_{i 1}(t),\ldots, x_{i p}(t))^T$ are predictors given as functions and 
$g_i \in$ $\{1,\ldots, L\}$ are classes to which each $x_i$ belongs.  
In the classification setting, we apply the Bayes rule, which assigns $x_i$ to class $g_i = l$ with the maximum posterior probability given $x_i$, denoted by ${\rm Pr}(g_i = l | x_i)$.  
Then the functional logistic regression model is given by the log-odds of the posterior probabilities:
\begin{align}
	\log \left \{ \frac {{\rm Pr}(g_i = l | x_i)}{{\rm Pr}(g_i = L | x_i)} \right \}
	= \beta_{0l} + \sum_{j=1}^p\int x_{i j}(t)\beta_{jl}(t)dt,
	\label{flogit1}
\end{align}
where $\beta_{0l}$ is an intercept and $\beta_{jl}(t)$ are coefficient functions.  
We assume that $x_{i j}(t)$ can be expressed by basis expansions as
\begin{align}
	x_{i j}(t) =
	\sum_{m=1}^{M_j} w_{i jm}\phi_{jm}(t) = 
	w^T_{i j}\phi_j(t),~~~
	\label{basis_x}
\end{align}
where 
$\phi_j(t)$ $=$ $(\phi_{j1}(t), \ldots, \phi_{jM_j}(t))^T$ 
are vectors of basis functions such as $B$-splines or radial basis functions, and 
$w_{i j}$ $=$ $(w_{i j1},\ldots, w_{i jM_j})^T$ are coefficient vectors.  
Since the data are originally observed at discrete time points, we smooth them with a basis expansion prior to obtaining the functional data $x_{i j}(t)$.  
In other words, $w_{i j}$ are obtained before constructing the functional logistic regression model (\ref{flogit1}).  
Details of the smoothing method are described in \cite{ArKoKa_etal2009a}.  
Furthermore, $\beta_{jl}(t)$ are also expressed by basis expansions 
\begin{align}
	\beta_{jl}(t) =
	\sum_{m=1}^{M_j} b_{jlm}\phi_{jm}(t) = 
	b^T_{jl}\phi_j(t), 
	\label{basis_beta}
\end{align}
where 
$b_{jl}$ $=$ $(b_{jl1}, \ldots, b_{jlM_j})^T$ 
are vectors of coefficient parameters.  

Using the notation
$\pi_l (x_i; b) = {\rm Pr}(g_i = l | x_i)$, where 
$b = (b_1^T,\ldots, b_p^T)^T$ and 
$b_j = (b_{j1}^T,\ldots, b_{j(L-1)}^T)^T$ 
since it is controlled by $b$, 
we can express the functional logistic regression model (\ref{flogit1}) as
\begin{align}
	\log \left \{ \frac {\pi_l ( x_i;  b)}{\pi_L ( x_i;  b)} \right \}
	= \beta_{0l} + \sum_{j=1}^p  w^T_{i j}\Phi_j b_{jl}
	=  \sum_{j=1}^p z_{ij}^T  b_{jl},
	\label{flogit2}
\end{align}
%where $ z_i = (1,  w^T_{i 1}\Phi_1, \ldots,  w^T_{i p}\Phi_p)^T$ and 
where $\Phi_j = \int \phi_j(t)\phi_j^T(t)dt$ and $z_{ij} = w_{ij}^T\Phi_j$.  
It follows from (\ref{flogit1}) that the posterior probability is 
\begin{align*}
	\pi_l( x_i;  b) &= \displaystyle{\frac {\exp \left ( \sum_j z_{ij}^T  b_{jl} \right )}
		{1 + \sum_{h=1}^{L-1} \exp \left ( \sum_j z_{ij}^T  b_{jh} \right )} \quad (l=1,\ldots,L-1)}, \\
	\pi_L( x_i;  b) &= \displaystyle{\frac {1} {1 + \sum_{h=1}^{L-1} \exp \left ( \sum_j z_{ij}^T  b_{jh} \right )}}.
\end{align*}
We define the vectors of the response variables $ y_i$, which indicate the class labels, as 
\begin{align}
	y_i = (y_{i 1},\ldots,y_{i (L-1)})^T = 
	\left\{
	\begin{array}{ll}
		(0,\ldots ,0,\stackrel{(l)}{1},0,\ldots,0)^T & {\rm if } \ g_i =l, ~~~l=1, \ldots, L-1, \notag \vspace{2mm}\\
		(0,\ldots,0)^T & {\rm if } \ g_i =L.
	\end{array}
	\right.
\end{align}
Then the functional logistic regression model has the probability function 
\begin{align}
	f( y_i |  x_i;  b) = \prod_{l=1}^{L-1} \pi_l( x_{i}; { b})^{y_{i l}} 
	\pi_L( x_{i}; { b})^{1 - \sum_{h=1}^{L-1} y_{i h}}.
	\label{logit-prob}
\end{align}

\section{Estimation by sparse regularization}
From the result of the previous section we can construct a likelihood function $\ell (b) = \sum_i \log f(x_i; b)$.  
This can be expressed as
\begin{align}
\ell ( b) = -\frac{1}{2}\left\|W^{1/2}\left({\eta} - \tilde Z b\right)\right\|_2^2,
\label{loglik}
\end{align}
where $W = (W_{hl})$ with
\begin{align*}
	W_{hl} = \left\{
	\begin{array}{ll}
	{\rm diag}\left\{\pi_{l}( x_1;  b)(1-\pi_{l}( x_1;  b)), \ldots, \pi_{l}( x_n;  b)(1-\pi_{l}( x_n;  b))\right\} & (h=l)  \\
	{\rm diag}\left\{-\pi_{h}( x_1;  b)\pi_{l}( x_1;  b), \ldots, -\pi_{h}( x_n;  b)\pi_{l}( x_n;  b)\right\} & (h\neq l), 
	\end{array}
	\right.
\end{align*}
and $W^{1/2}$ is a matrix that satisfies $W = W^{1/2}W^{1/2}$.  
Furthermore, $\tilde Z = (\tilde Z_1,\ldots, \tilde Z_p)$ with $\tilde Z_j = I_{L-1}\otimes Z_j$ and $Z_j = (z_{1j}, \ldots, z_{nj})^T$, ${\eta} = \tilde Z b + W^{-1}\Lambda 1_{n(L-1)}$, 
$\Lambda = {\rm diag}\left\{\Lambda_1, \ldots, \Lambda_{L-1}\right\}$, 
$\Lambda_l = {\rm diag}\left\{ y_{1l}-\pi_{l}( x_1;  b),\ldots, y_{nl}-\pi_{l}( x_n;  b)\right\}$, and 
$1_{n(L-1)} = (1,\ldots, 1)^T$ is an $n(L-1)$-dimensional vector.
Then we consider maximizing the penalized log-likelihood function
\begin{align}
\ell_{\lambda, \alpha} (b) = \ell (b) - P_{\lambda,\alpha}(b),
\label{pls1}
\end{align}
where we assume the sparse group lasso-type penalty for $P_{\lambda,\alpha}(b)$:
\begin{align}
P_{\lambda,\alpha}(b) = n(1-\alpha)\sum_{j=1}^p \lambda_j\left\{\sum_{l=1}^{L-1}\|b_{jl}\|_2^2\right\}^{1/2} - 
n\alpha\sum_{j=1}^p \lambda_j\sum_{l=1}^{L-1}\|b_{jl}\|_2, 
\label{pen}
\end{align}
where $\lambda_j = \sqrt{M_j}\lambda$ with a regularization parameter $\lambda>0$ and $\alpha \in [0,1]$ is a tuning parameter.  
The first term of this penalty has an effect that it shrinks some of $b_j$ towards exactly zero vectors, using the idea of the group lasso, and it leads to variable selection, while the second term shrinks $b_{jl}$ toward zero vectors separately, which leads to decesion boundary selection.

We want to estimate $b$ by maximizing the function (\ref{pls1}), but there are two difficulties in deriving the estimator.  
First, it is generally difficult to explicitly express the parameters estimated by the sparse regularization.  
In order to solve this problem we use the idea of the coordinate descent alogirhtm \citep{FrHaHo_etal2007}.   
Second, when we apply the sparse group lasso-type penalty it is difficult to construct updated values for parameters if the design matrices for each of the groups (in our case $\tilde Z_1, \ldots, \tilde Z_p$) are not orthogonal \citep{FrHaTi2010b}.  
\cite{SiFrHa2013} approached this problem by applying the Taylor expansion and the majorization-minimization algorithm.  
On the other hand, we apply the QR decomposition by using the idea of \cite{SiTi2012} to form the orthogonal design matrix.  
The QR decomposition provides $W^{1/2}\tilde Z_j = Q_j R_j$, where $Q_j$ is an orthogonal matrix and $R_j$ is an upper triangle matrix.  
Denote $b_j^* = R_j b_j$, then the log-likelihood function (\ref{loglik}) can be re-expressed by 
\begin{align*}
\ell(b^*) = -\frac{1}{2}\left\|W_j^{1/2}r_{-j} - Q_j b_j^*\right\|_2^2,
\end{align*}
where $r_{-j} = \eta - \sum_{j'\neq j} Z_{j'} b_{j'}$.

The partial derivative of $\ell_{\lambda,\alpha} (b^*)$ with respect to $b_j^*$ is given by
\begin{align*}
\frac{\partial \ell_{\lambda} (b^*)}{\partial b_j^*} = 
\tilde r_{-j} - b_j^{*} - n(1-\alpha)\lambda_j u_j - n\alpha\lambda_j v_j,
\end{align*}
where $\tilde r_{-j} = (\tilde r_{-j1}, \ldots, \tilde r_{-j(L-1)})^T = Q_j^TW_j^{1/2}r_{-j}$ and $u_j$ and $v_j = (v_{j1}, \ldots, v_{j(L-1)})^T$ are vectors of subgradients respectively given by
\begin{align*}
u_j = \left\{
\begin{array}{ll}
\displaystyle{\frac{b_j^*}{\|b_j^*\|_2}} & (b_j^* \neq 0) \\
{\rm s.t.~~} \|u_j\|_2 \ge 1 & (b_j^* = 0), 
\end{array}
\right. \\
v_{jl} = \left\{
\begin{array}{ll}
\displaystyle{\frac{b_{jl}^*}{\|b_{jl}^*\|_2}} & (b_{jl}^* \neq 0) \\
{\rm s.t.~~} \|v_{jl}\|_2 \ge 1 & (b_{jl}^* = 0).  
\end{array}
\right.
\end{align*}

Let $ S_j = (S_{j1},\ldots, S_{j(L-1)})^T$ with $S_{jl} = (\|\tilde r_{jl}\|_2 - n\alpha\lambda)_+$ be vectors of thresholding functions, where $(a)_+ = \max\{a, 0\}$, then if $\| S_{j}\|_2 \le n(1-\alpha)\lambda$ the parameter vector $b_j^*$ is estimated to be $\hat b_{j}^*=0$.  
Otherwise, solve the following equation with respect to $b_{jl}^*$: 
\begin{align*}
\frac{\partial \ell_{\lambda} (b^*)}{\partial b_{jl}^*} = 
\tilde r_{-jl} - b_{jl}^{*} - n(1-\alpha)\lambda_j \frac{b_{jl}^{*}}{\|b_{j}^{*}\|_2} - n\alpha\lambda_j v_{jl} = 0.
\end{align*}
Then, if $\|\tilde r_{jl}\|_2\le n\alpha\lambda$ then $\hat b_{jl}^* = 0$, otherwise $\hat b_{jl}^*$ is calculated as
\begin{align*}
\hat b_{jl}^* = 
\frac{\|\hat b_{j}^*\|_2(\|\tilde{r}_{-jl}\|_2-n\alpha\lambda)_+}
     {\|\hat b_{j}^*\|_2+n(1-\alpha)\lambda}
\frac{\tilde{ r}_{-jl}}{\|\tilde{r}_{-jl}\|_2},
\end{align*}
where $\|\hat b_{j}^*\|_2$ is given by
\begin{align*}
\|\hat b_{j}^*\|_2 &= \left(\| h_j\| - n(1-\alpha)\lambda\right)_+, \\
h_j &= ( h_{j1}^T,\ldots,  h_{j(L-1)}^T)^T, 
h_{jl} = \left(\|\tilde{r}_{-jl}\|_2-n\alpha\lambda\right)_+
\frac{\tilde{r}_{-jl}}{\|\tilde{r}_{-jl}\|_2}.
\end{align*}

The algorithm is given in the following steps:
\begin{enumerate}
	\item (Outer loop) For $j=1,\ldots, p$, check if $\|S_{j}\|_2 \le n(1-\alpha)\lambda$.  
	If it is true, $\hat b_j^* = 0$ and if not, for each $j$ go to Step 2.  
	\item (Inner loop) For $l=1,\ldots, L-1$, update $ b_{jl}^*$ as follows:
\begin{align*}
\hat b_{jl}^* = 
\frac{\|\hat b_{j}^*\|_2(\|\tilde{r}_{-jl}\|_2-n\alpha\lambda)_+}
{\|\hat b_{j}^*\|_2+n(1-\alpha)\lambda}
\frac{\tilde{r}_{-jl}}{\|\tilde{r}_{-jl}\|_2}.
\end{align*}
	\item Iterate Step 1 and 2 until convergence and then obtain estimators $\hat b_1^*,\ldots, \hat b_p^*$.  
	\item Calculate $\hat b_j = R_j^{-1}b_j^*$ for each $j$.  
\end{enumerate}
The outer loop corresponds to the variable selection step, and the inner loop corresponds to the decision boundary selection step.

The statistical model estimated by the above method strongly depends on tuning parameters $\lambda$ and $\alpha$.  
In order to decide appropriate values for them, we apply a model selection criterion.  
Although the cross validation is commonly used for selecting such parameters, it needs multiple computations for estimation and may often be computationally expensive.  
On the other hand, various criteria based on information criterion or Bayesian information criterion (BIC) are used to evaluate models from viewpoints of prediction accuracy and model selection consistency.  
Here we apply a BIC-type model selection criterion.  
\cite{ZhLiTs2010} showed that the BIC-type criterion consistently select models when we apply the SCAD penalty \citep{FaLi2001}.  
The effective degrees of freedom is obtained by the trace of the (pseudo) smoother matrix.  
The smoother matrix of our model is obtained by calculating
\begin{align*}
	\tilde Z_j\hat b_{j}
%	&= \tilde Z_j R_j^{-1}\hat b_{j}^*\\
	&= \tilde Z_j R_j^{-1} C_j Q_j^TW_j^{1/2}r_{-j} \\
	&= S_j r_{-j},
\end{align*}
where $S_j = \tilde Z_j R_j^{-1} C_j Q_j^TW_j^{1/2}$ and $C_j$ is given by
\begin{align*}
	C_j = \left(\begin{array}{ccc}
	c_{j1} 1_{M_j}^T & \cdots & 0 \\
	\vdots & \ddots & \vdots \\
	0 & \cdots & c_{j(L-1)} 1_{M_j}^T
	\end{array}\right),\\
	c_{jl} = 
	\frac{\|\hat b_{j}^*\|_2(\|\tilde{r}_{-jl}\|_2-n\alpha\lambda)_+}
	{\|\hat b_{j}^*\|_2+n(1-\alpha)\lambda}
	\frac{1}{\|\tilde{r}_{-jl}\|_2}.
\end{align*}
We consider $S_j$ as a smoother matrix of our model, and therefore the effective degrees of freedom is given by ${df = \sum_j{\rm tr}S_j}I(\|\hat b_j\|_2 \neq 0)$.
Thus we have a model selection criterion
\begin{align*}
{\rm BIC} = -2\ell(\hat{b}) + {df}\log n.
\end{align*}
We choose the values of $\lambda$ and $\alpha$ that minimize BIC and then regard the corresponding model as an optimal model.  
\section{Example with real data }
\begin{figure}[t]
	\begin{tabular}{cc}
		\begin{minipage}{0.33\hsize}
			\begin{center}
				\includegraphics[height=4.5cm,width=4.5cm]{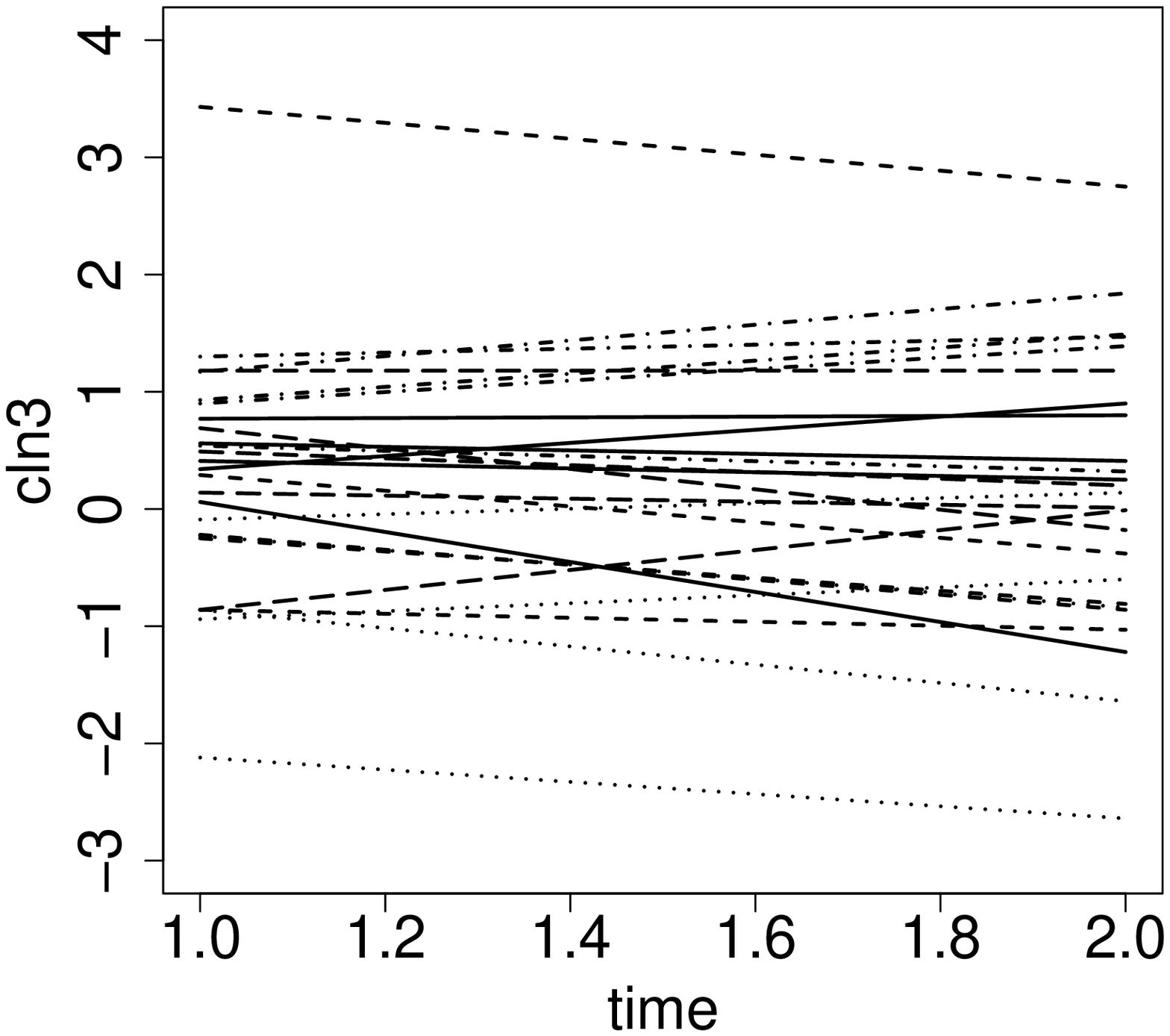}
			\end{center}
		\end{minipage}
		\begin{minipage}{0.33\hsize}
			\begin{center}
				\includegraphics*[height=4.5cm,width=4.5cm]{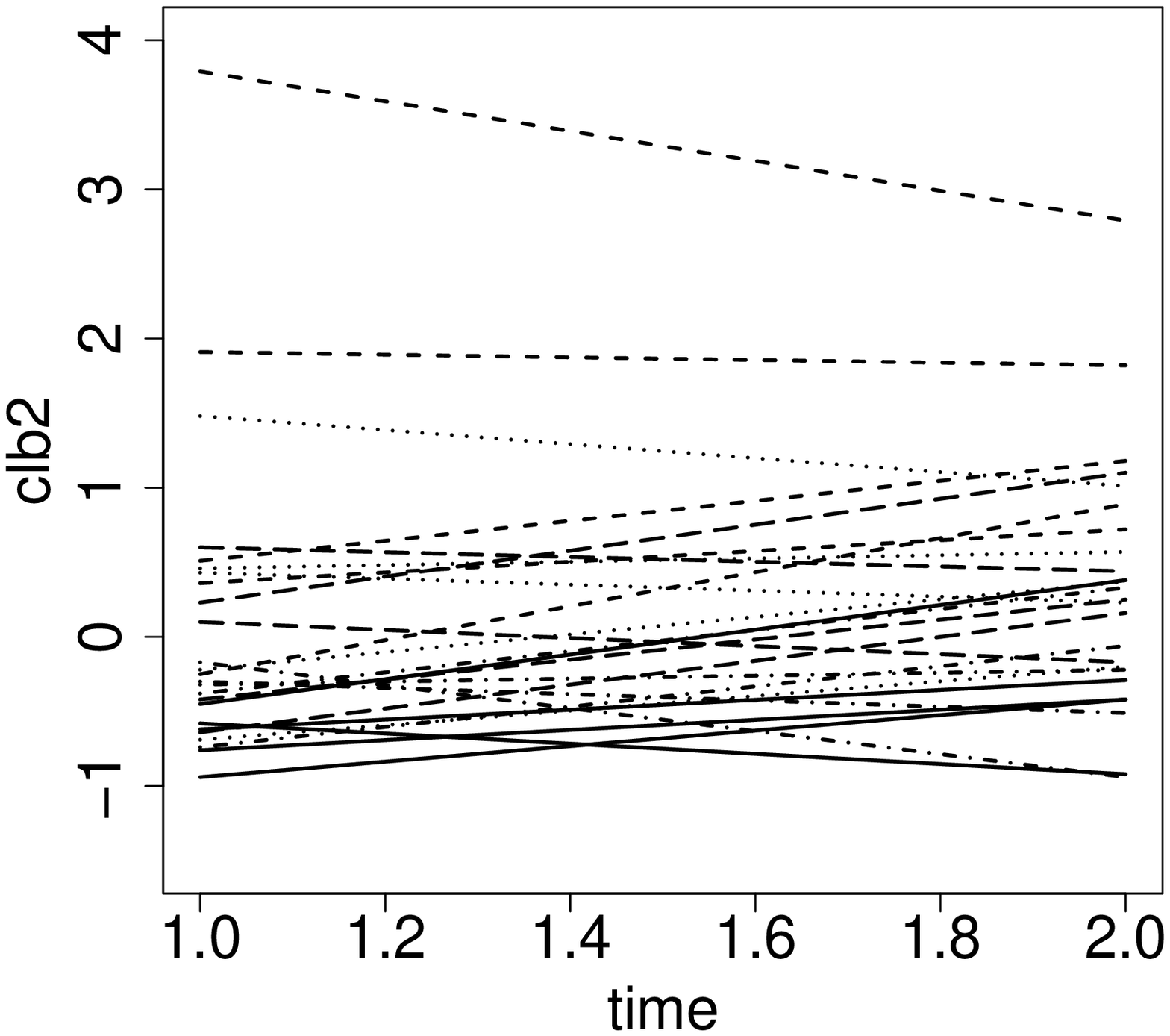}
			\end{center}
		\end{minipage}
		\begin{minipage}{0.33\hsize}
			\begin{center}
				\includegraphics*[height=4.5cm,width=4.5cm]{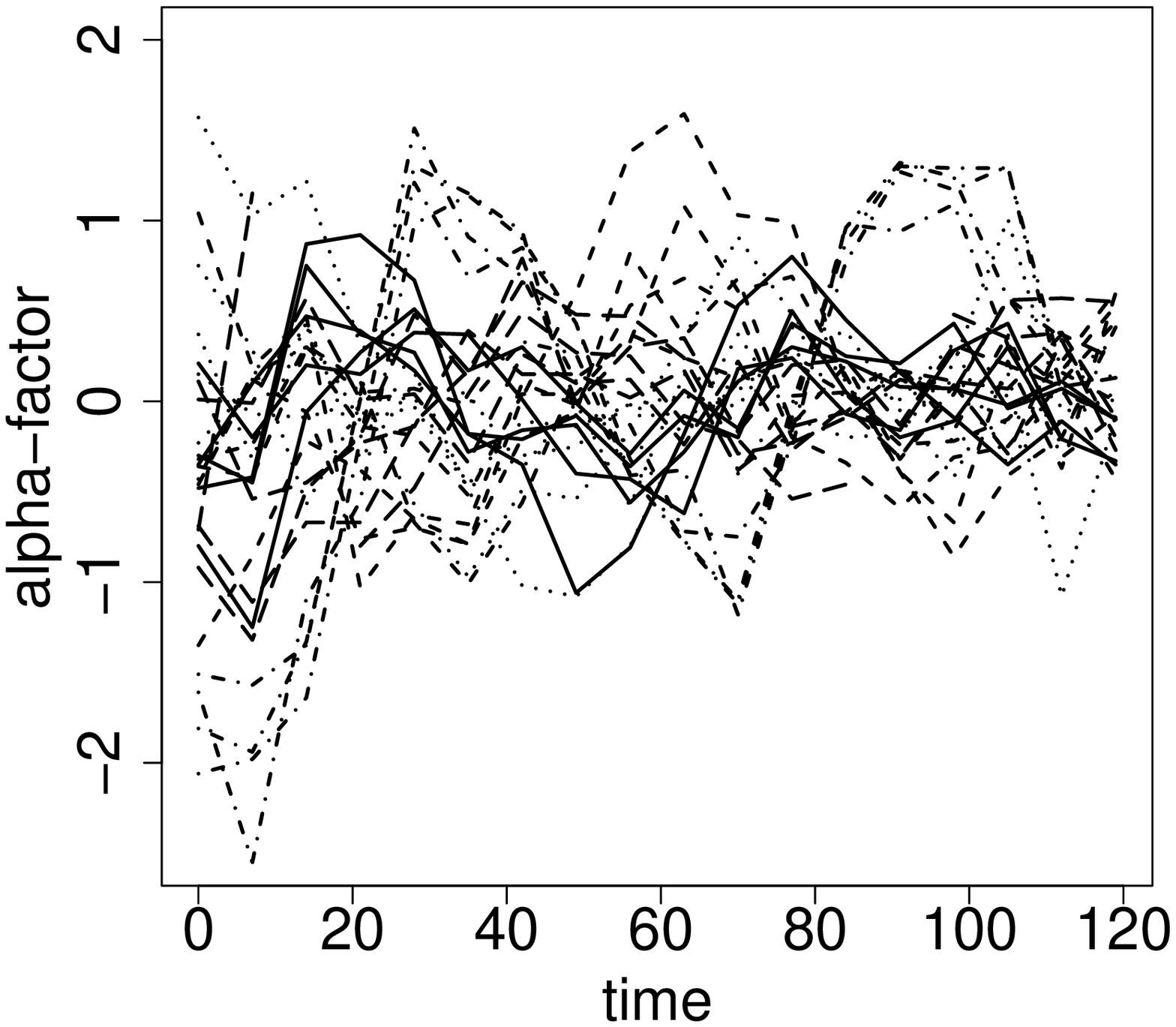}
			\end{center}
		\end{minipage}\\
	\end{tabular}
	\begin{tabular}{cc}
		\begin{minipage}{0.33\hsize}
			\begin{center}
				\includegraphics[height=4.5cm,width=4.5cm]{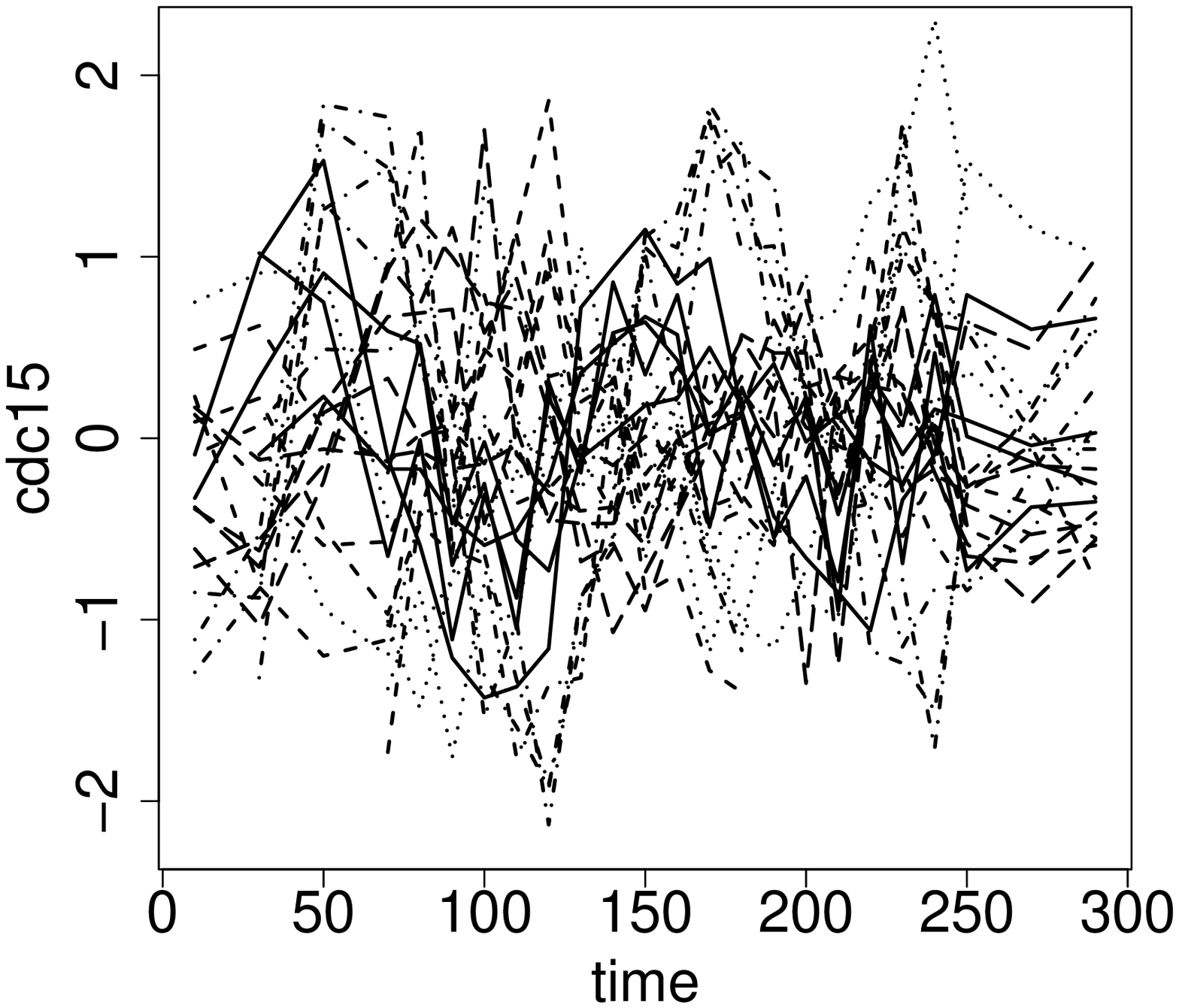}
			\end{center}
		\end{minipage}
		\begin{minipage}{0.33\hsize}
			\begin{center}
				\includegraphics*[height=4.5cm,width=4.5cm]{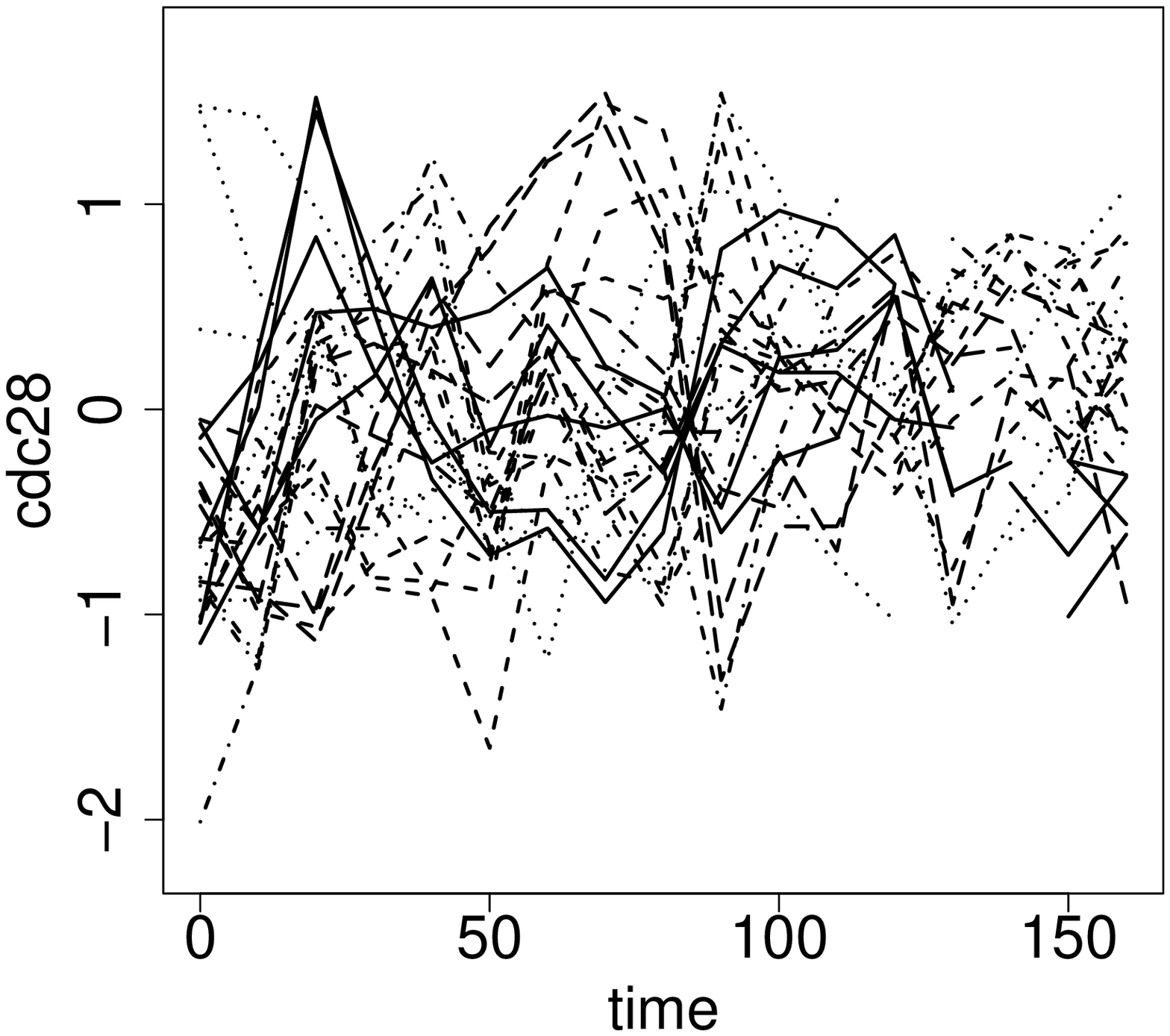}
			\end{center}
		\end{minipage}
		\begin{minipage}{0.33\hsize}
			\begin{center}
				\includegraphics*[height=4.5cm,width=4.5cm]{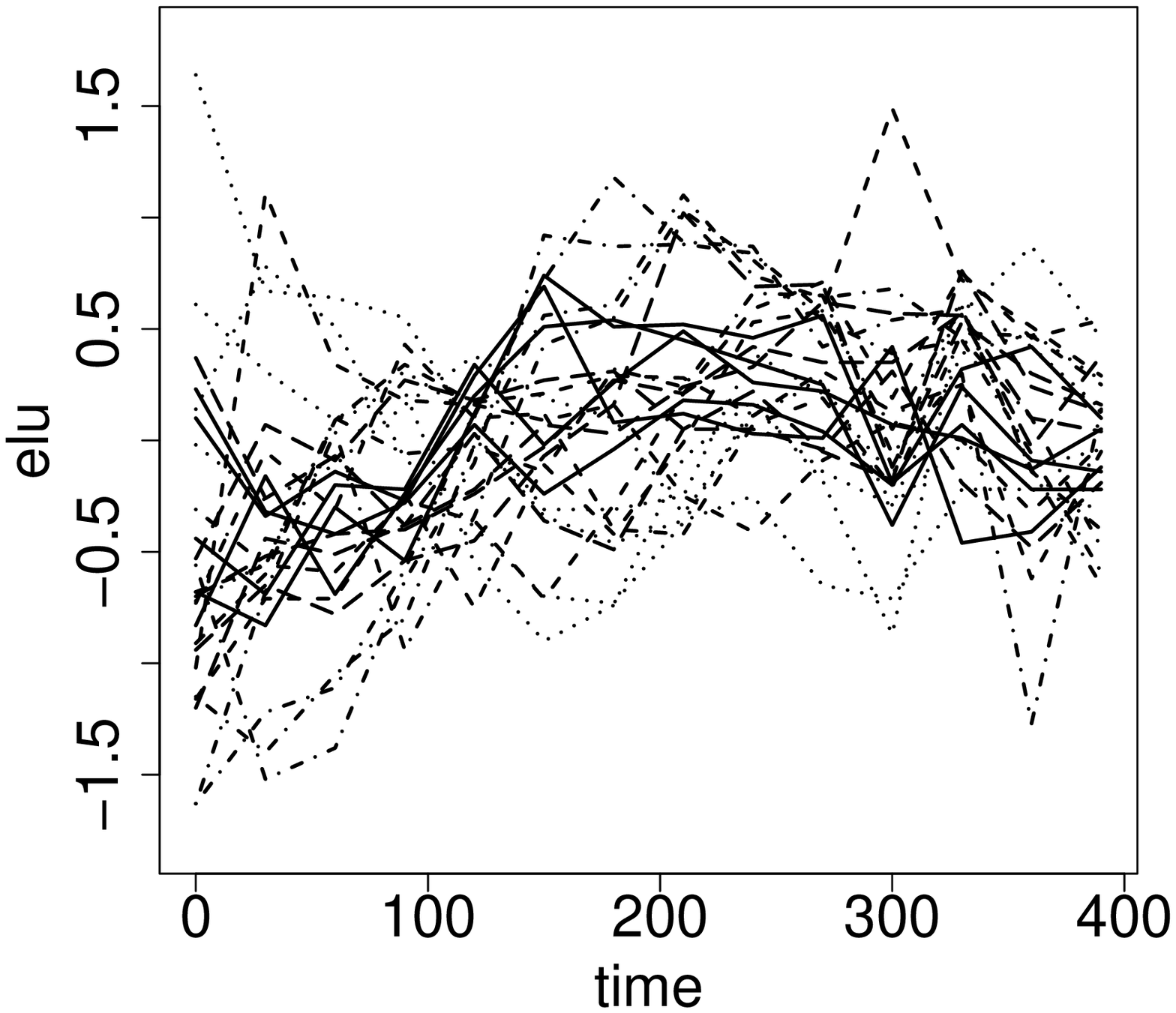}
			\end{center}
		\end{minipage}
	\end{tabular}
	\caption{
		Yeast cell cycle gene expression profiles for each type of synchronization.  Each plot consists of 5 genes from 5 classes:  G1 (solid), G2/M (dashed), M/G1 (dotted), S (dot-dashed), and S/G2 (long dashed).
	}
	\label{fig:yeast}
\end{figure}

We applied the proposed method to the analysis of yeast cell cycle gene expression data.  
\cite{SpShZh_etal1998} measured expression profiles over about two cell cycles for 6,178 genome-wide yeast genes using cDNA microarrays.  
The data contain 77 microarrays with several types of temporal synchronization: cln3 (2 points), clb2 (2 points), $\alpha$-factor (18 points), cdc15 (24 points), cdc28 (17 points), and elu (14 points).  
\cite{SpShZh_etal1998} used the clustering method from the above 77 experiments to classify 800 genes into 5 groups: G1, G2/M, M/G1, S, and S/G2.  
Figure \ref{fig:yeast} shows examples for each type of synchronization.  
\cite{ArKoKa_etal2009b} classified genes by using the cdc15 experiments as functional data and then used the posterior probabilities to determine the misclassified data.  
Here we consider if these 6 experiments affect the classification.  

Since there are many missing values in the expression profiles and only 72 genes have no missing values, we excluded genes according to the following two rules:  
(1) Genes with at least one missing value for either cln3 or clb2 were excluded.  
(2) Those with a total of more than 10 missing values from some combination of $\alpha$-factor, cdc15, cdc28, and elu were excluded.  
We can easily apply the regression model even if there are some (not excessively many) missing values by converting them into functional data.  
The resulting 657 genes were used for this analysis.  
First, except for cln3 and clb2, we smoothed the time-course data to construct functions.  
They were expressed using basis expansions with 4 basis functions that were previously selected.  
The remaining variables, cln3 and clb2, each of which have only 2 time points, were treated as vector data rather than functional data.  
We also treated the variables corresponding to the 2 time points as a group.  
Then we constructed a functional logistic regression model as follows:
\begin{align}
\log \left \{ \frac {{\rm Pr}(g_i = l |  x_i)}{{\rm Pr}(g_i = L |  x_i)} \right \}
= \beta_{0l} + 
\sum_{j=1}^2\sum_{j'=1}^2 x_{i j_{j'}}\beta_{j_{j'}l} + 
\sum_{j=3}^6\int x_{i j}(t)\beta_{jl}(t)dt,
\label{flogit_yeast}
\end{align}
which is a special case of (\ref{flogit1}), where $x_{ij}$ $(j=1,\ldots, 6)$ correspond to gene expression profiles for cln3, clb2, $\alpha$-factor, cdc15, cdc28, and elu, respectively.  
The model was estimated by the penalized likelihood method with the sparse group lasso-type penalty and then the regularization parameter was selected by the BIC.  
We repeated this process for 50 bootstrap samples.  
Furthermore, we altered the class label $L$ on the left-hand side of (\ref{flogit_yeast}) and repeatedly estimated the model in order to investigate all the coefficients of the decision boundaries.  
As a result, there are totally 100 repetitions for the model for all combinations of two classes.  
We then investigated which variables and decision boundaries affected the classification.  

Table \ref{tab:yeast} shows the numbers of selected decision boundaries for bootstrap samples.  
We found that many coefficients were estimated to be nonzero.  
However, the coefficient for the boundary between M/G1 and S/G2 was not selected at all for clb2, and, similarly, those between M/G1 and S and S and S/G2 were rarely selected.  
This indicates that the variable clb2 does not affect the above classifications.  
On the other hand, Table \ref{tab:yeast2} shows that all the variables are selected for each of the 100 repetitions in the viewpoint of variable selection.  
This result indicates that all of the variables themselves are relevant to the classification.  

%1:G1  2:G2/M  3:M/G1  4:S  5:S/G2
%cln3, clb2, alpha, cdc15, cdc28, elu
\begin{table}[t]
	\caption{
		Numbers of selected decision boundaries. 
	}
	\label{tab:yeast}
	\begin{center}
		%\begin{tabular}{c|C{.05}C{.05}C{.05}|C{.05}C{.05}C{.05}}
		\begin{tabular}{ccccccc}
			\hline\hline
			& cln3 & clb2 & $\alpha$ & cdc15 & cdc28 & elu \\
			\hline
			G1 vs. G2/M   & 100 & 100 & 91 & 100 & 100& 64\\
			G1 vs. M/G1   & 100 & 60  & 98 & 70  & 95 & 95\\
			G1 vs. S      & 88  & 64  & 94 & 99  & 97 & 100\\
			G1 vs. S/G2   & 92  & 60  & 99 & 99  & 100& 98\\
			G2/M vs. M/G1 & 46  & 52  & 55 & 100 & 90 & 45\\
			G2/M vs. S    & 90  & 55  & 34 & 99  & 93 & 52\\
			G2/M vs. S/G2 & 75  & 51  & 52 & 100 & 70 & 45\\
			M/G1 vs. S    & 63  & 9   & 55 & 79  & 71 & 69\\
			M/G1 vs. S/G2 & 48  & 0   & 100& 100 & 94 & 79\\
			S vs. S/G2    & 29  & 2   & 51 & 55  & 48 & 51 \\
			\hline
		\end{tabular}
	\end{center}
	\caption{
	Numbers of selected variables. 
}
\label{tab:yeast2}
\begin{center}
	%\begin{tabular}{c|C{.05}C{.05}C{.05}|C{.05}C{.05}C{.05}}
	\begin{tabular}{cccccc}
		\hline\hline
		 cln3 & clb2 & $\alpha$ & cdc15 & cdc28 & elu \\
		\hline
		    100 & 100 & 100 & 100 & 100& 100\\
		\hline
	\end{tabular}
\end{center}
\end{table}

\section{Concluding remarks}
We have proposed the method for selecting both variables and decision boundaries in estimating the multiclass logistic regression model for functional data.  
We derived the estimation and evaluation procedures for the model with the sparse group lasso-type penalty.  
The model was fitted by the penalized maximum likelihood method using the blockwise coordinate descent algorithm, and then the tuning parameters involved in the model was selected by the model selection criterion.  
The sparse group lasso penalty is composed of two terms; the group lasso and the lasso.  
The former has a role of selecting variables, on the other hand, the latter selects decision boundaries.  
%Monte Carlo simulations were conducted in order to investigate the effects on the accuracy of prediction and on variable selection.  
The proposed method was applied to the analysis of gene expression data, and we then investigated which types of time synchronization contributed to the classification of cell cycles.  

For estimating the model with the sparse group lasso penalty, several algorithms are proposed such as \cite{SiFrHa2013} and \cite{ViHa2014}.  
Furthermore, the alternating direction method of multipliers by \cite{BoPaCh_etal2011} appears to be useful for estimating our model.  
We will consider the application and comparison of these methods as future works.  
We derived the BIC using the idea of the effective degrees of freedom, but the BIC is originally derived from the framework of the maximum likelihood method.  
The derivation of the model selection criterion for our model estimated by the penalized likelihood method is also a future work.   
\section*{Acknowledgments}
This work was supported by Grant-in-Aid for Young Scientists (B) No. 25730017 and No. 16K16020 of JSPS.

\end{document}